# Optical Anisotropy in Bismuth Titanate: An Experimental and Theoretical Study

**Amritendu Roy [1], Rajendra Prasad [2], Sushil Auluck [3] and Ashish Garg [1†]**

[1] Department of Materials Science and Engineering, Indian Institute of Technology Kanpur; Kanpur- 208016; India
[2] Department of Physics, Indian Institute of Technology Kanpur; Kanpur- 208016; India
[3] National Physical Laboratory, K.S. Krishnan Marg. New Delhi, India

## ABSTRACT

We report experimental and theoretical investigation of anisotropy in optical properties and their origin in the ferroelectric and paraelectric phases of bismuth titanate. Room temperature ellipsometric measurements performed on pulsed laser deposited bismuth titanate thin films of different orientations show anisotropy in the dielectric and optical contstants, Subsequent first-principles calculations performed on the ground state structures of ferroelectric and high temperature paraelectric phases of bismuth titanate show that the material demonstrates anisotropic optical behavior in both ferroelectric and paraelectric phases. We further show that O 2p to Ti 3d transition is the primary origin of optical activity of the material while optical anisotropy results from the asymmetrically oriented Ti-O bonds in $TiO_6$ octehdra in the unit cell.

**Keywords:** Bismuth titanate, ferroelectric, optical properties, ellipsometry, first-principles calculations

## I. INTRODUCTION

Dielectric, optical and optoelectronic properties of bismuth titanate ($Bi_4Ti_3O_{12}$ or, BiT) are of particular interest in the area of optical memories since its band gap, similar to most ferroelectric perovskite oxides, lies in the visible spectrum region. As a result, dielectric [1-9] and optical properties [9-15] of BiT have been extensively investigated for decades. In addition, there has been a renewed interest in the optical properties of ferroelectric BiT as noncentrosymmetric crystal structure of room temperature BiT could trigger asymmetric electron excitation, relaxation and scattering leading to photovoltaic effect.[16, 17]

Ferroelectric BiT with layered perovskite structure has a complex crystal structure consisting of alternate stacking of fluorite $(Bi_2O_2)^{2+}$ and perovskite-like $(Bi_2Ti_3O_7)^{2-}$ layers arranged along the crystallographic *c*-axis. Noncentrosymmetry of the room temperature structure along with its complexity further renders anisotropic physical properties in the material, namely dielectric constant ($\varepsilon_a = \varepsilon_b = 153 \pm 5$ and $\varepsilon_c = 118 \pm 5$ at 100 kHz)[2], ferroelectric polarization,[11, 18] thermal[19] and electrical[20, 21] conductivities. It is interesting to note that anisotropy in different

[†] Corresponding author. Tel: +91-512-2597904; Fax: +91-512-2597505; email: ashishg@iitk.ac.in

physical properties have different origins. For example, directional dependence of thermal properties is attributed to the density difference between the perovskite and the fluorite type ($Bi_2O_2$) layers in the unit cell *[19]* while anisotropy in the electrical properties arises due to differences in the oxygen vacancy concentrations in the perovskite and bismuth oxide layer of the structure.*[20, 21]* In contrast, anisotropy in the ferroelectric polarization is attributed to non-zero displacement of Ti ions along *a*-axis.*[18]*

From the perspective of optical devices utilizing oriented thin films of BiT, it is vital to experimentally and theoretically explore the optical anisotropy if any and then to understand its origin. Here, first-principles density functional calculations would be useful since such study could shed a detailed insight into the origin of optical activity, as shown previously for a number of oxides.*[16, 22-24]* Though considerable attention has been paid toward the experimental studies of optical and dielectric properties of BiT*[11, 13, 15]*, very few reports provide a microscopic understanding of the optical properties using first-principles calculations. Among very few reports, Cai *et al.[25]* based on the bond orbital theory concluded that the large nonlinear refractive index of ferroelectric BiT is due to the virtual electronic excitation from the filled valence band to the empty cationic *d*-orbital at short equilibrium bond lengths. Recently, optical properties of ferroelectric bismuth titanate have been calculated using TB-mBJ functional using a smaller cell with space group *Pc [26]* where close proximity between experimental and calculated band gap was found however, orientation dependence of optical behavior was not discussed.

In this manuscript, we show the results of ellipsometric measurements on epitaxial bismuth titanate thin films of two different orientations grown on (100) and (110) oriented strontium titanate (STO) substrates using pulsed laser deposition. Optical constants determined by fitting the ellipsometric data demonstrate direction dependence pointing towards optical anisotropy in the material. Further, we calculated the optical properties in ferroelectric and paraelectric phases of bismuth titanate using first principles density functional theory based calculations using local density approximation (LDA) and generalized gradient approximation (GGA) methods as well as full-potential based LAPW method using GGA technique. The calculations show orientation dependence of dielectric constant not only in ferroelectric phase, but also in the paraelectric phases of BiT. We also show that the optical activity in both the phases originates primarily from O 2p to Ti 3d transition.

## II. EXPERIMENTAL DETAILS AND CALCULATION METHODS

Thin films of BiT were grown using pulsed laser deposition technique on strontium titanate substrates with two different orientations *viz.* (100) and (110) from a stoichiometric target of BiT. The film growth was carried out using an excimer laser of wavelength 248 nm (KrF) in an oxygen environment ($pO_2 \sim 0.40$ mbar) at a substrate temperature of 750°C using a laser fluence of 2.3 J/cm$^2$ and a repetition rate of 5 Hz. Prior to deposition, a base pressure of $2\times10^{-6}$ mbar was achieved. The films were annealed at the deposition temperature in the oxygen environment with $pO_2 \sim 0.67$ mbar for 30 min. and then cooled to room temperature at the deposition pressure. X-ray diffraction of the films was performed using a high resolution Philips X'Pert PRO MRD thin film diffractometer using CuKα radiation over 2θ range of 10° to 80°. Ellipsometric measurements were carried out using HORIBA JOBIN-YVON spectroscopic ellipsometer (SE) over the energy range of 0.8-2.5 eV with an incidence angle of 70°. In the present work, we have used a three layer model in which complex dielectric function and other optical properties were determined by simulating the experimental data using Tauc-Lorentz (TL) model.*[27]* Separate ellipsometric measurements were performed on the bare substrates in order to separate out their contributions.

We started our first-principles calculations with the optimized lattice parameters and ionic positions of *B1a1* and *I4/mmm* phases of bismuth titanate calculated using GGA and LDA. *[18]* The entire calculation was carried out in the framework of first-principles density functional theory *[28]*. Vienna ab-initio simulation package (VASP) *[29, 30]* was used with projector augmented wave method (PAW).*[31]* The Kohn-Sham equation *[32, 33]* was solved using the exchange correlation function of Perdew and Wang *[34, 35]* for generalized gradient approximation (GGA) and of Ceperley-Alder *[36]* for local density approximation (LDA) schemes. We included five valence electrons for Bi ($6s^2 6p^3$), 4 for Ti ($3d^3 4s^1$) and 6 for O ($2s^2 2p^4$). A plane wave energy cut-off of 400 eV was used. Conjugate gradient algorithm *[37]* was used for the structural optimization. All the calculations were performed at 0 K. For electronic structure and optical property calculation, Monkhorst-Pack *[38]* 8×8×8 mesh was used. In order to substantiate our results using pseudopotential based approach, we repeated our entire calculations by FP-LAPW method with GGA using WIEN2k. Such a comparison would be first of its kind for a complex system such as bismuth titanate.

### III. RESULTS AND DISCUSSIONS

**A. Optical Characterization using Ellipsometry**

Fig. 1(a) plots XRD spectra of single phase BiT thin films deposited on two differently oriented STO substrates. The bottom panel depicts (001)-oriented BiT with presence of (001) peaks. The top panel shows the film with (118)-orientation with major peak being (2 2 16) peak ((118) reflection is not seen due to systematic absences). A weaker (117) peak is also observed which is quite close to the calculated position of (118) peak. On the other hand, (001) and (118) are ~ 47° apart from each other. Subsequently, we performed ellipsometric measurements on these two differently oriented films to study the effect of orientation on linear optical responses. Fig. 1(b) shows the three layer model used for fitting the ellipsometric data using TL model.*[27]* Fig. 1(c) shows plots of experimental and fitted data for (001) oriented BiT film suggesting the validity of the model used in the present study. Calculated film thickness is consistent with the profilometer data (~700-750 nm). Obtained band gap is also consistent with previous reports. *[39]*

Subsequently, real and imaginary components of dielectric spectra obtained from the fitting were plotted for the two films and are shown in Fig, 2(a). Since both films have approximately identical thicknesses, optical response can also expected to be similar provided the material was isotropic. In contrast, dielectric spectra of the two films, as shown in Fig. 2(a), demonstrate different intensities translating into different optical constants for two films. Inset of Fig. 2(a) plots absorption coefficients (α) as a function of energy for the two films. For semiconductors, absorption coefficient follows $E^{1/2}$ and $E^2$ relationships for direct and indirect gap semiconductors, respectively. We find that the absorption spectra beyond the band gap of both the films follow eq. (1) nicely as shown below:

$$\alpha = \alpha_0 \left( \frac{E - E_g}{E_g} \right)^2 \quad \ldots\ldots\ldots\ldots\ldots\ldots\ldots\ldots\ldots (1)$$

where $E_g$ represents the band gap. This substantiates that BiT is an indirect band gap semiconductor, in agreement with our previous work[18] and several other experimental[15, 39] and first-principles calculations. [26] Upon fitting our absorption data using eq. (1), we obtained energy band gap of the order of $E_g$ = 3.57±0.01 eV and $E_g$ = 3.54±0.02 eV for (001) and (118) oriented films, respectively which are consistent with previously reported data.[39] Below the band gap, our samples manifest phonon assisted Urbach absorption which can be conveniently modeled by:

$$\alpha = \alpha_g \left( \frac{E - E_g}{E_U} \right) \quad\quad\quad\quad\quad (2)$$

where $E_U$ indicates the Urbach absorption energy. Use of eq. 2 in the low absorption Urbach region yield $E_g$ = 3.64±0.02 eV for (001) oriented film. Subsequent calculations of optical constants such as refractive index (*n*), extinction coefficient (*k*) and reflectivity (*R*) for the two differently oriented films are presented in Fig.2(b) and (c). These figures show different optical behaviors for the above two films again pointing towards optical anisotropy in BiT. For instance, refractive indices for the two films when extrapolated to zero frequencies are 2.44 and 2.27, respectively for (001) and (118) orientations. These values are consistent with previous experiments.[39] Reflectivity and optical conductivity too demonstrate different magnitudes when plotted as a function of incident energy.

In order to develop an in-depth understanding of the observed optical response in ferroelectric phase of BiT we further performed first-principles calculations on the ground state structure of BiT in its ferroelectric state. In addition, we also carried out first-principles calculations for optical properties on the high temperature paraelectric phase of BiT.

**B First-principles Calculations and Origin of Optical Anisotropy**
In the ferroelectric *B1a1* symmetry of BiT, the real and imaginary parts of the dielectric constant tensor would have three unequal leading diagonal elements, $\varepsilon_{xx}$, $\varepsilon_{yy}$ and $\varepsilon_{zz}$ along with off-diagonal terms due to monoclinic symmetry. Since, the monoclinic distortion is very small (β = 90.08°), we ignore the off-diagonal elements and present only the diagonal elements. Paraelectric BiT, on the other hand, has tetragonal (*I4/mmm*) symmetry and therefore, there would be two components of real and imaginary parts of dielectric constant:

$$\left. \begin{array}{l} \varepsilon_\perp = \dfrac{\varepsilon_{xx} + \varepsilon_{yy}}{2} \\ \varepsilon_\parallel = \varepsilon_{zz} \\ \varepsilon_{avg.} = \dfrac{\varepsilon_\parallel + 2\varepsilon_\perp}{3} \end{array} \right\} \quad\quad\quad\quad (3)$$

Fig 3 (a)-(f) show the calculated real and imaginary parts of dielectric constants plotted as a function of incident photon energy for *B1a1* and *I4/mmm* phases using GGA and LDA schemes calculated using pseudopotential and full-potential approaches. We do not observe any significant difference between the spectra calculated using pseudopotential and full-potential approaches. This supports the robustness of our calculation. Calculated spectra for GGA and LDA are also qualitatively identical with certain differences in the intricate details. We therefore, limit our discussion within the GGA results since the same would be applicable to LDA as well.

For the ferroelectric phase, a comparison between the calculated and experimental dielectric spectra reveals that (i) there is difference in position of the major features of the spectra and (ii) there is intensity difference as well. While the difference in intensity could arise from a number of factors such as nature and quality of sample, temperature, difference between experimental and ground state structural parameters, type of approximation scheme used in the first-principles calculations and type of broadening used in the experimental and calculated data, the difference in peak position is chiefly attributed to the underestimation of energy band gap by LDA and GGA techniques, an inherent drawback of such calculations. If the calculated spectra are shifted to higher energy by 1 eV, we find good agreement between experimental data and the calculations. Thus, our further analysis would ignore the underestimation of band gap and corresponding mismatch of the peak positions with respect to the experimental spectra and would concentrate more on the origin of the optical spectra. Our calculation on the ferroelectric phase also reveals that there are numerous peaks in the imaginary part of the dielectric constant versus energy plot. In order to identify the peaks, we, fitted multiple Lorentzian functions and labeled them. The goodness of fitting for all cases is > 0.99. Table 1 lists the position of the peaks of the xx component of $\varepsilon''$ computed using GGA. Fig 3(a) shows that $\varepsilon''$ spectra for three principal components have peaks at different energy values attributed to different optical transitions allowed at those energies. The anisotropy in the intensity of dielectric spectra, in Fig. 3(a), results in different optical constants along principal crystallographic directions (not shown here). For instance, refractive, indices, extinction coefficients, reflectivity and real parts of optical conductivities along three principal crystallographic directions demonstrate that while xx and zz components of the above optical constants are similar, the magnitudes of yy component differs significantly in *B1a1* symmetry of BiT. This result is consistent with the work of Singh et al*[26]* where a co-ordinate transformation to *B1a1* symmetry allows the comparison.

In a similar manner, we also calculated dielectric spectra of the paraelectric *I4/mmm* phase. Tetragonal symmetry of the system allows only two components of the dielectric spectra, viz., $\varepsilon_\perp$ and $\varepsilon_\parallel$ as defined in eq. 3. Fig. 3(d) shows real and imaginary components of dielectric constant for tetragonal *I4/mmm* symmetry calculated using GGA. Comparison of both real and imaginary components of $\varepsilon_\perp$ and $\varepsilon_\parallel$ depicts that the high symmetry phase of BiT demonstrates significant anisotropic behavior in optical properties which is further substantiated by the anisotropic optical constants (not shown here). Similar to the ferroelectric phase, we identified prominent peaks in the imaginary $\varepsilon_\parallel$ spectra and listed them in Table 1.

To identify the origin of these peaks we subsequently performed electronic structure calculations on the ground state structures of the above two phases. Fig 4(a) and (b) show the parts of the band structures along with the transitions that resulted the above described optical activities in ferroelectric *B1a1* and paraelectric *I4/mmm* BiT, respectively. Fig. 4(c) and (d) shows corresponding site projected density of states. The calculations were done using GGA. Corresponding LDA plots are similar with minute differences and therefore not shown here. Detailed discussions on the density of states and band structures of the two phases can be found elsewhere. *[18]* The uppermost part of the valence band (VB) consists mainly of O *2p* states. Above the Fermi level, the conduction band (CB) is dominated by Ti *3d* states where contribution from Ti2 ion more significant than that of Ti1. Noticeable amount of Bi 6p and O 2p states are also present here. There, we propose that major optical transition would involve O 2p, Bi 6p and Ti 3d states.

Crystal structure of BiT consists of alternate perovskite and fluorite layers and the density of $TiO_6$ octahedra is different in crystallographic *c*-direction with respect to the other two directions. Such anisotropic structural feature has been reported to result in anisotropic physical properties such as electrical conductivity.[20] Although, the density of $TiO_6$ octahedra is identical along *a* and *b* directions, the orientation of $TiO_6$ octahedra is different in those two directions. Such orientation difference of Ti-O bonds results in preferential development of spontaneous polarization along crystallographic *a* direction and not in *b* direction in BiT.[18] Further, we notice that, in *I4/mmm* structure, optical responses are identical in *a* and *b* direction with symmetrical Ti-O bonds. Thus, it is plausible that the orientation differences among the Ti-O bonds lead to different transition behavior of O 2p to Ti 3d, primarily responsible for optical activity of BiT, rendering anisotropic optical behavior in BiT.

## IV. CONCLUSIONS

To summarize, we performed a combined experimental-theoretical study on the optical properties of ferroelectric and paraelectric phases of bismuth titanate. Our work shows that bismuth titanate demonstrates anisotropic optical properties in both ferroelectric as well as paraelectric phases. Major optical activity is attributed to the O2p-Ti3d electronic transition while optical anisotropy is attributed to the asymmetrically oriented Ti-O bonds in the unit cell.

**Acknowledgement**

The work was supported by Department of Science and Technology, Govt. of India through project number SR/S2/CMP-0098/2010. Authors thank Surajit Sarkar for his assistance in ellipsometry measurement. SA thanks CSIR National Physical Laboratory, New Delhi for financial assistance.

Table 1  Peak positions and peak labeling in ε″ spectra of *B1a1* and *I4/mmm* structures.

| *B1a1* | (eV) | *I4/mmm* | (eV) |
|---|---|---|---|
| B$_1$ | 2.93 | I$_1$ | 3.21 |
| B$_2$ | 3.38 | I$_2$ | 3.55 |
| B$_3$ | 3.97 | I$_3$ | 4.07 |
| B$_4$ | 4.6 | I$_4$ | 4.51 |
| B$_5$ | 5.04 | I$_5$ | 5.03 |
| B$_6$ | 5.41 | I$_6$ | 5.38 |
| B$_7$ | 5.9 | I$_7$ | 5.83 |

**Figure Captions:**

Fig. 1 (a) XRD of BiT thin films with orientations (001) (lower panel) and (118) (upper panel) deposited on SrTiO$_3$ single crystalline substrates with orientations (100) and (110), respectively. (b) Three layer model used for simulating the experimental ellipsometry data. Above the substrate, L1 represent an interface between substrate and BiT film, L2 shows BiT thin film and L3 indicates top surface of the film with some roughness. (c) Simulation of ellipsometry data of (001) oriented BiT film using Tauc-Lorez model, shown the validity of the present model.

Fig. 2 optical properties of BiT thin films with (001) and (118) orientations obtained from ellipsometric measurements. Inset plots absorption co-efficient as a function of incident photon energy.

Fig.3 Real and imaginary parts of dielectric constant plotted as a function of incident photon energy for ferroelectric B1a1 ((a), (b) and (c)) and paraelectric I4/mmm ((d), (d) and (f)) phases calculated using GGA ((a) and (d)) and LDA ((b) and (e)) methods within vasp and (c) and (f) calculated using GGA within WIEN2k.

Fig.4 Partial DOS and band structures of (a) ferroelectric (*B1a1*) and (b) paraelectric (*I4/mmm*) bismuth titanate calculated using GGA.

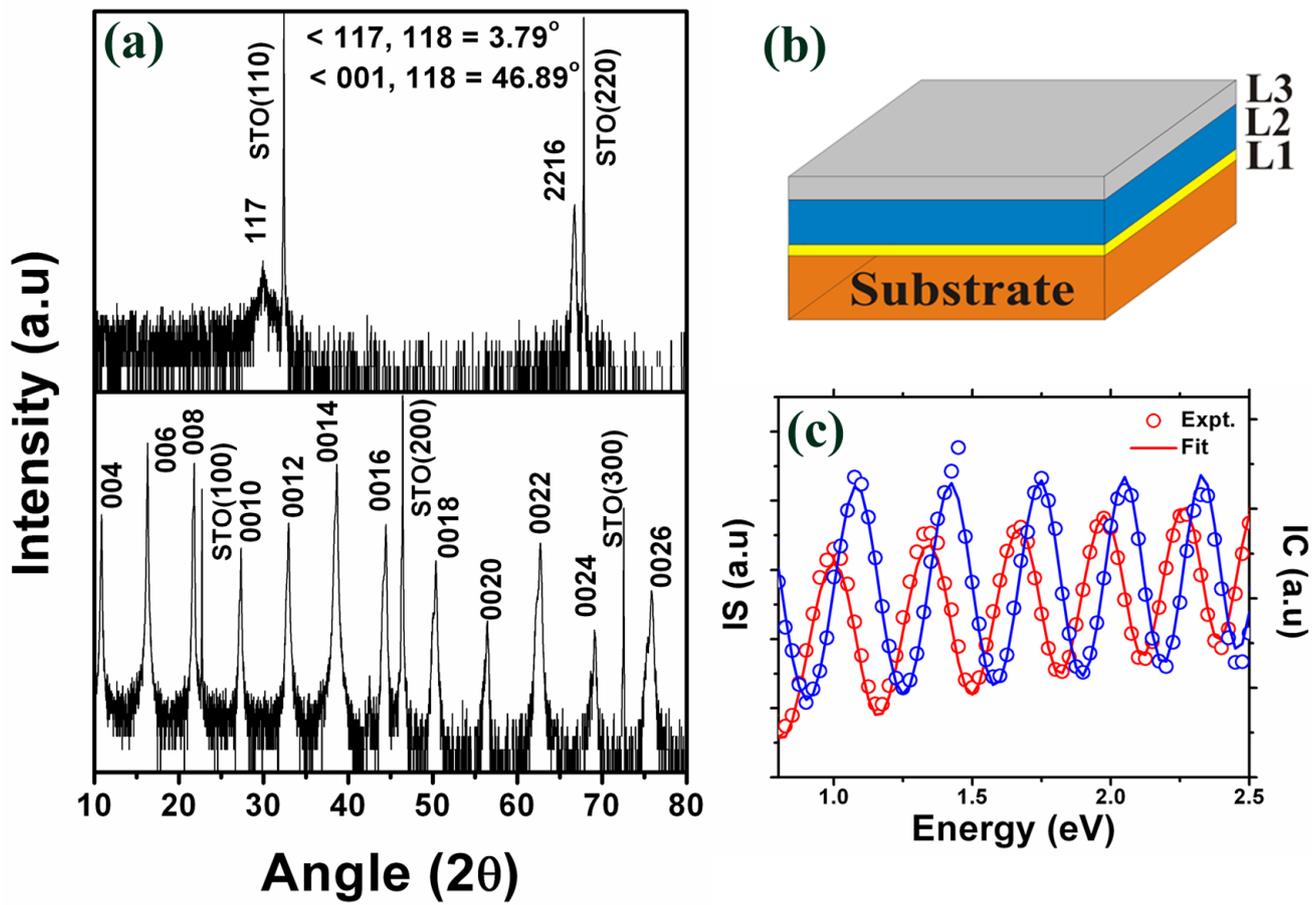

Figure 1 (Fig.1.tif)

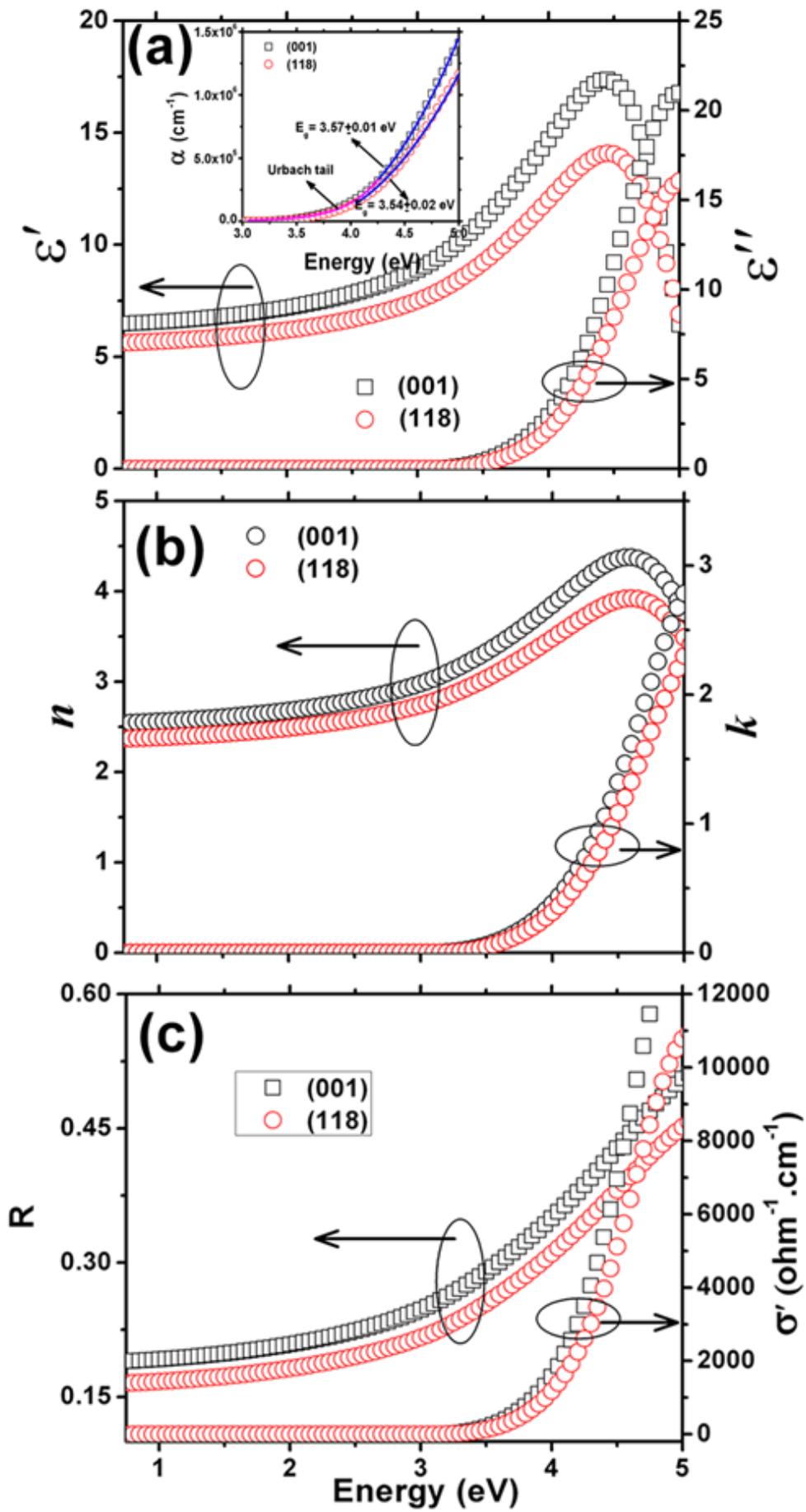

**Figure 2 (Fig.2.tif)**

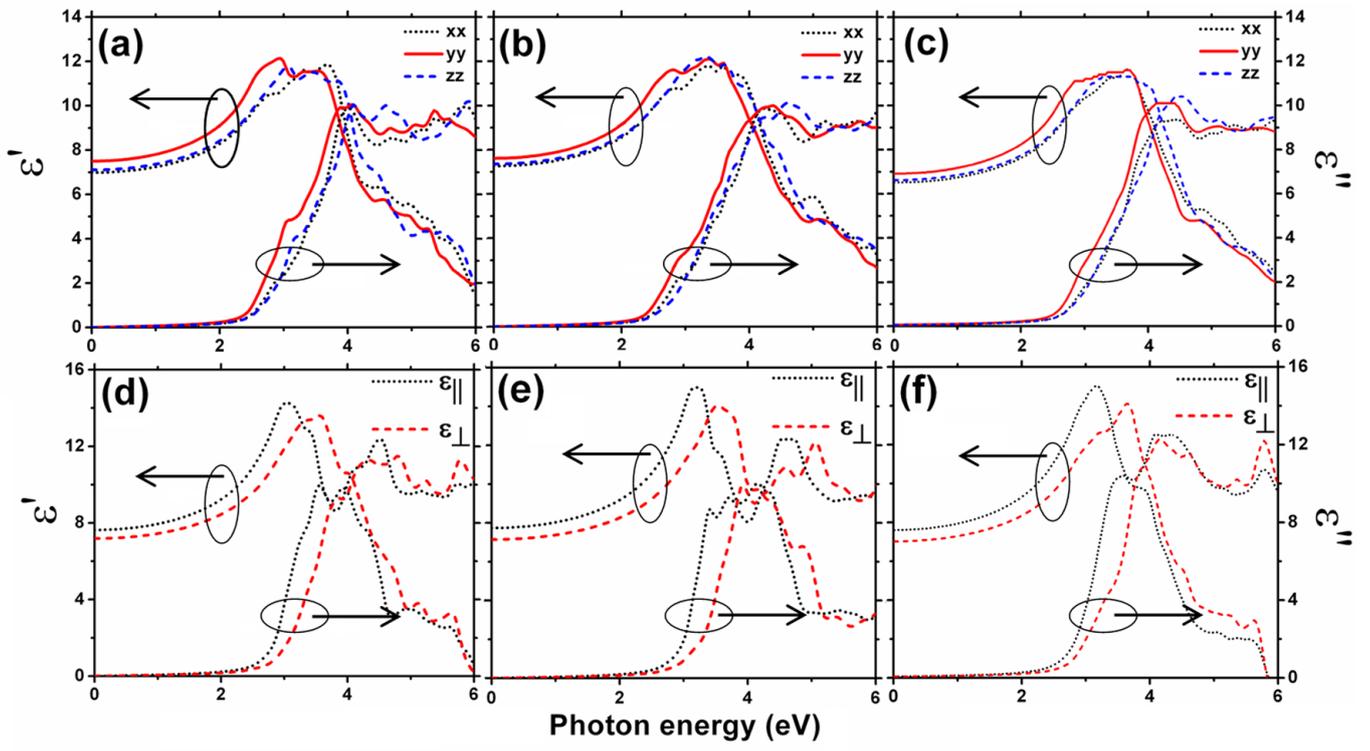

**Figure 3 (Fig.3.tif)**

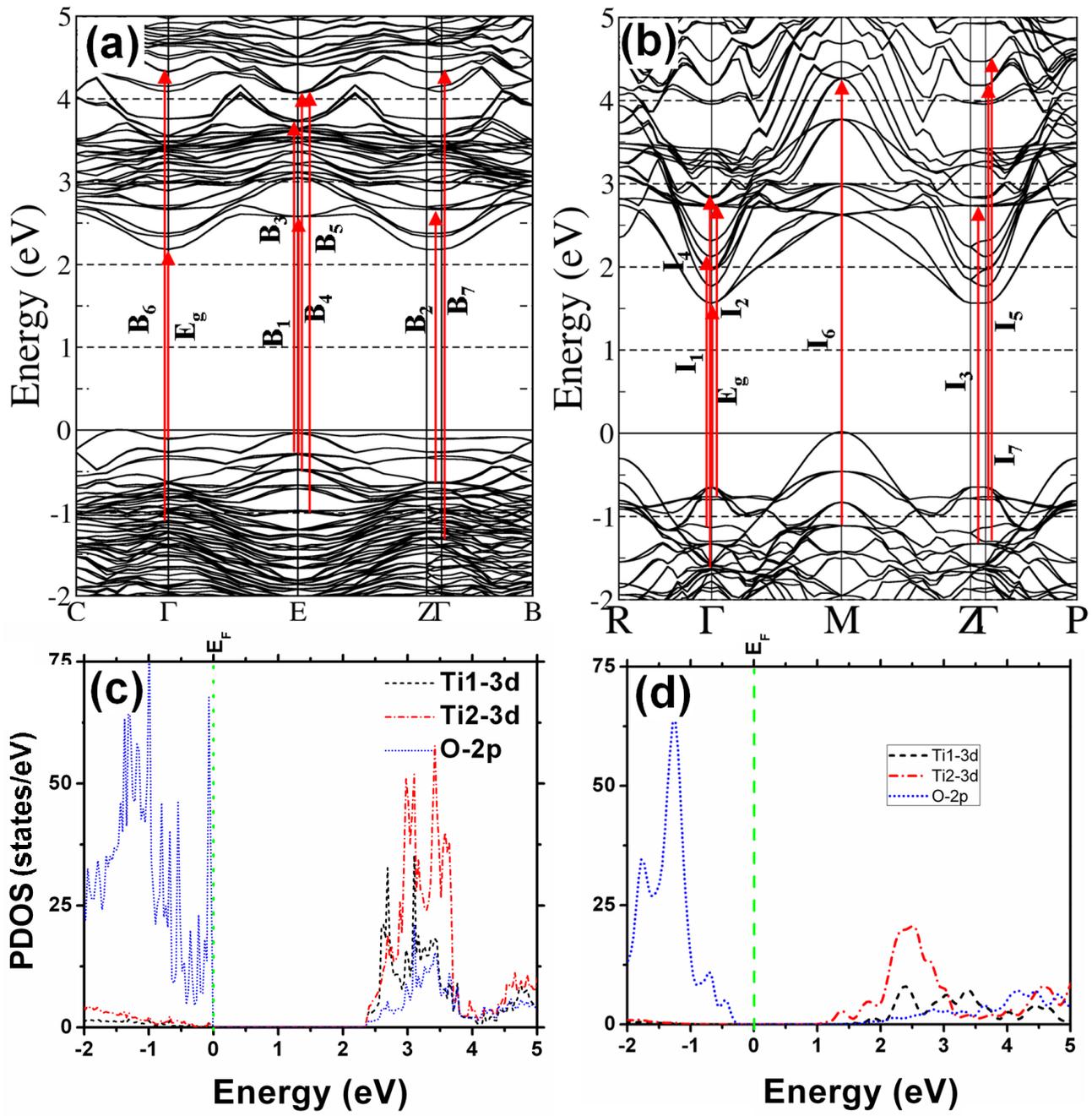

Figure 4 (Fig. 4.tif)